\documentclass{article}

\usepackage{natbib}
\usepackage{amssymb,amsfonts,amsbsy}

\newtheorem{theorem}{Theorem}
\newtheorem{lemma}{Lemma}
\newtheorem{proposition}{Proposition}
\newtheorem{problem}{Problem}
\newtheorem{definition}{Definition}
\newtheorem{corollary}{Corollary}
\newtheorem{remark}{Remark}
\newtheorem{example}{Example}
\newtheorem{algo}{Algorithm}

\newtheorem{conjecture}{Conjecture}
\newtheorem{diagram}{Diagram}
\newtheorem{notation}{Notation}

\newenvironment{proof}{\noindent\textit{Proof:}}{\hfill{\rule{0.5em}{1.2ex}}\\}
\newenvironment{analysis}{\noindent\textit{Analysis:}}{\hfill{\rule{0.5em}{1.2ex}}\\}

\newcommand{\K}{\mathbb{K}}
\newcommand{\F}{\mathbb{F}}
\newcommand{\E}{\mathbb{E}}
\newcommand{\LL}{\mathbb{L}}
\newcommand{\N}{\mathbb{N}}
\newcommand{\Q}{\mathbb{Q}}
\newcommand{\Z}{\mathbb{Z}}
\newcommand{\D}{\mathbb{D}}
\newcommand{\x}{\mathbf{x}}
\newcommand{\y}{\mathbf{y}}
\newcommand{\z}{\mathbf{z}}
\newcommand{\td}{\mathrm{tr.deg.}}
\newcommand{\Ker}{\mathrm{Ker}}
\newcommand{\rank}{\mathrm{rank}}
\newcommand{\car}{{\rm char}}
\newcommand{\Nor}{{\rm N}}
\newcommand{\Res}{{\rm Res}}
\newcommand{\Aut}{{\rm Aut}}

\begin{document}

\title{Computation of unirational fields}

\author{\small Jaime Gutierrez\footnote{Both authors are partially supported by Spain's Ministerio de Educaci\'on Grant
Project BFM2001-1294}\\\small Dpto. de Matem\'aticas, Estad\'{\i}stica y
Computaci\'on, Univ. de Cantabria\\\small E--39071 Santander, Spain\\\small
jaime.gutierrez@unican.es \and \small David Sevilla\\\small Department of
Computer Science, Concordia University\\\small Montreal H3G 1M8, QC,
Canada\\\small dsevilla@cs.concordia.ca}

\date{}

\maketitle

\begin{abstract}
One of the main contributions which Volker Weispfenning  made to mathematics
is related to Gr\"obner bases theory. In this paper we present an algorithm
for computing all algebraic intermediate subfields in a separably generated
unirational field extension (which in particular includes the zero
characteristic case). One of the main tools is Gr\"obner bases theory.
 Our algorithm also requires  computing
primitive elements and factoring over algebraic extensions. Moreover, the
method can be extended to finitely generated $\K$-algebras.
\end{abstract}

\section{Introduction}

The goal of this paper is to study the problem of computing
intermediate fields between a rational function field and a given
subfield of it. Rational function fields arise in various contexts
within mathematics and computer science. Two examples are the
%%%DAVID "algebraic geometric" --> "algebraic geometry" , "reparametrizing" --> reparametrization
factorization of regular maps in algebraic geometry \citep{Sha77}
and the reparametrization of parametric varieties in computer aided
geometric design \citep{AGR99}.

The question of the structure of the lattice of such intermediate
fields is of theoretical interest by itself; we will focus on the
computational aspects, like deciding if there are proper
intermediate fields and computing them in the affirmative case.

In the univariate case, the problem can be stated as follows: given
an arbitrary field $\K$ and $f_1,\ldots,f_m\in\K(t)$, find a field
$\F$ such that
$\K(f_1,\ldots,f_m)\varsubsetneq\F\varsubsetneq\K(t)$. By L\"uroth's
Theorem, see \citep{Wae64}, or \citep{Sch82} for a constructive
proof by Netto, there exist functions $f,h\in\K(t)$ such that
$\K(f_1,\ldots,f_m)=\K(f)$ and $\F=\K(h)$. Therefore, our problem is
equivalent to decomposing the rational function $f$, that is, to
find $g,h\in\K(t)$ with $\deg\,g$, $\deg\,h>1$ such that $f=g(h)$.
Algorithms for decomposition of univariate rational functions can be
found in \citep{Zip91} and \citep{AGR95}.

We denote by $\K$ an arbitrary field and by
$\K(x_1,\ldots,x_n)=\K(\x)$  the rational function field in the
variables $\x=(x_1,\ldots,x_n)$. In the multivariate case, the
problem can be stated as:

\begin{problem}\label{problem-1}
Given rational functions $f_1,\ldots,f_m\in\K(\x)$, compute a proper
unirational field $\F$ between $\K(f_1,\ldots,f_m)$ and $\K(\x)$, if
it exists.
\end{problem}

%%%DAVID "unirational" resaltado
A \emph{unirational} field over $\K$ is an intermediate field $\F$
between $\K$ and $\K(\x)$.  We know that any unirational field is
finitely generated over $\K$, see \citep{Nag93}. Thus, by computing
an intermediate field we mean that such a finite set of generators
is to be calculated.

Regarding algorithms for this problem, see \citep{MS99}, where the
authors generalize the method of \citep{AGR95} to several variables,
by converting this problem into the calculation of a primary ideal
decomposition. Primary ideal decomposition can be computed by
Gr\"obner bases. The book \citep{BW93} is an excellent reference
guide to this important theory and its applications. Once the
primary ideal decomposition is computed in a polynomial ring with
$2n$ variables, their algorithm requires to check a exponential
number of generators of the possible intermediate proper subfields
--- although authors do not study its complexity in detail. On the
other hand, the solution is trivial and uninteresting for most
choices of $f_1,\ldots,f_m$, since it is easy to construct
infinitely many intermediate fields when the transcendence degree of
$\K(f_1,\ldots,f_m)$ over $\K$ is smaller than $n$, as the next
theorem shows.

\begin{theorem}
If $n>\td(\K(f_1,\ldots,f_m)/\K)$, there exist infinitely many
different fields between $\K(f_1,\ldots,f_m)$ and $\K(\x)$.
\end{theorem}

\begin{proof}
At least one of $x_1,\ldots,x_n$ is transcendental over
$\K(f_1,\ldots,f_m)$, let us assume that $x_1$ is. Then the fields
\[\K(f_1,\ldots,f_m,x_1^k)\ ,\quad k\in\N\]
form an infinite set of different intermediate fields. Indeed, if
$i$ divides $j$,
\[\K(f_1,\ldots,f_m,x_1^j)\varsubsetneq\K(f_1,\ldots,f_m,x_1^i).\]

It is clear that one field is contained in the other. To prove that
they are not equal, assume that $x_1^i\in\K(f_1,\ldots,f_m,x_1^j)$.
Then there exists a rational function $h(t)$ such that
$x_1^i=h(x_1^j)$ where $h\in\K(f_1,\ldots,f_m,t)$ but
$h\not\in\K(t)$. Then we have the polynomial relation
\[x_1^i\cdot h_D(f_1,\ldots,f_m,x_1^j)-h_N(f_1,\ldots,f_m,x_1^j)=0,\]
(where $h_N$, $h_D$ denote the numerator and denominator of $h$
resp.) which contradicts $x_1$ being transcendental over
$\K(f_1,\ldots,f_m)$.
\end{proof}

Due to this result, we will focus on the following version of the
problem.

\begin{problem}\label{prob-alg}
Given functions $f_1,\ldots,f_m \in\K(\x)$, find all the fields $\F$
between $\K(f_1,\ldots,f_m)$ and $\K(\x)$ that are algebraic over
$\K(f_1,\ldots,f_m)$.
\end{problem}

%%%DAVID "separated basis" --> "separating basis"
First, we will prove that there are finitely many algebraic
intermediate fields if the original extension is separable. The
notion of separable extension can be generalized to non-algebraic
extensions. In transcendental extensions, separability means that
any finitely generated subfield $\F$ over $\K $ has a separating
basis, that is, a transcendence basis $B$ such that $\K(B)\subset
\F$ is an algebraic separable extension. The following is a well
known result, see for instance \citep{Lan67}.

\begin{proposition}~\label{cor-sep}
The field extension $\K\subset \K(\x)$ is separable.
\end{proposition}

In general, if $\K'$ is a separable extension of $\K$, then every
field between $\K$ and $\K'$ is separable over $\K$. Details on
separability and a proof of these results can be found in
\cite{Nag93} and \cite{Lan67}.

As we said, any unirational field is finitely generated over $\K$.
The following result provides a bound for the number of generators
and it is known for zero characteristic field. Our algorithm always
returns this bound as the number of generators.

\begin{theorem}\label{cota-gens}
Let $\F$ be a unirational field such that
$\K\varsubsetneq\F\subset\K(\x)$ and $d=\td(\F/\K)$. Then there
exist $h_1,\ldots,h_s\in\K(\x)$ such that $\F=\K(h_1,\ldots,h_s)$
and $s\leq d+1$.
\end{theorem}

\begin{proof}
By Proposition~\ref{cor-sep} we have $\K\subset \K(\x)$  is
separable, that is, for each subfield $\F$ in $\K\subset \K(\x) $
there exists a transcendence basis $\{h_1,\dots,h_d\}$ of $\F$ over
$\K $ such that $\K(h_1,\dots,h_d)\subset \F $ is algebraic
separable. Then, the result follows by the Primitive Element
Theorem.
\end{proof}

Because of the previous results we have the following theorem.

\begin{theorem}
If the extension $\K(\x)/\K(f_1,\ldots,f_m)$ is separable then there exist finitely many intermediate fields that are
algebraic over $\K(f_1,\ldots,f_m)$.
\end{theorem}

\begin{proof}
Let $\F_0$ be the minimum subfield of $\K(\x)$ that contains all algebraic intermediate fields. $\F_0$ is clearly
algebraic over $\K(f_1,\ldots,f_m)$, and due to the previous theorem the extension $\F_0/\K(f_1,\ldots,f_m)$ is
separable. On the other hand, since $\F_0$ is a unirational field  is finitely generated over $\K$, see Theorem
\ref{cota-gens}. Therefore, because of the Primitive Element Theorem the extension is simple and there are finitely
many fields between $\K(f_1,\ldots,f_m)$ and $\F_0$.
\end{proof}

Problem \ref{prob-alg} for transcendence degree of $\K(f_1,\ldots,f_m)/\K$ is 1 has been treated in \citep{GRS01}. In
this case a generalization of the classical L\"uroth's Theorem applies:
\\

\noindent \textbf{Extended L\"uroth's Theorem } Let $\F$ be a field
such that $\K\subset\F\subset\K(x_1,\ldots,x_n)$ and $\td(\F/\K)=1$.
Then there exists $f\in\K(x_1,\ldots,x_n)$ such that $\F=\K(f)$.
Also, if the field contains a polynomial, then a polynomial
generator exists.
\\

%%%DAVID Cambiado x1,...,xn por \x en casi todo el paper, empezando aqui
By the Extended L\"uroth's Theorem, the problem is equivalent to the following: given $f\in\K(\x)$, find $g\in\K(y)$
and $h\in\K(\x)$ with $\deg\,g$, $\deg\,h>1$ such that $f=g(h)$. The paper \citep{GRS02} provides a very efficient
constructive proof of the above result and it also contains different decomposition algorithms for multivariate
rational functions. In some sense, Problem \ref{prob-alg} can be seen as a generalization of the univariate rational
function decomposition problem.

In this paper we will combine several techniques of computational algebra to create an algorithm that finds all the
intermediate fields that are algebraic over the smaller field. Moreover, our method can be extended to finitely
generated $\K$-algebras, that is, the case where the ambient field is $\K(z_1,\ldots,z_n)=\K(\z)$ for some
$z_1,\ldots,z_n$ transcendental over $\K$ (that need not be algebraically independent), and $\K(\z)$ is the quotient
field of a polynomial ring, so that we have
\[\K(\z)=QF\left(\K[\x]/I\right)\]
(where $QF$ denotes the quotient field of a domain) for some prime ideal $I\subset\K[\x]$ that will be given
explicitly by means of a finite system of generators. Unsurprisingly, the algorithm will be much simpler when $\K(\x)$
is rational, that is, when $I=(0)$.

The paper is organized as follows. In Section 2 we introduce several algebraic tools in order to manipulate fields and
the elements in them. Section 3 is devoted to the algebraic case, and in Section 4 the general case is reduced to it,
also other approaches to this are given. Section 5 briefly describes the adaptation of the algorithm to $\K$-algebras.
Finally, in Section 6 we summarize the main conclusions of this research and consider some open problems.

\section{Algebraic tools}

In this section we will introduce several techniques and tools of
general interest for the manipulation of fields and functions.

\emph{Notation.} Through this paper, we will denote the numerator and denominator of a rational function $f$ as $f_N$
and $f_D$ respectively.

\subsection{Membership problem}\label{subsect-sweedler}

As we have to manipulate function fields and field extensions, we
may need to compute generators and elements with certain properties,
or to check whether certain functions belong to a given field. The
next theorem provides a way to do this. See \citep{Swe93} and for
more details, see also \citep{BW93}.

We will use the following notation: Let $A$ be a commutative
$\K$-algebra and $\{a_0,\ldots,a_n\}$ be a set of generators of $A$
over $\K$. Let $\K[x_0,\ldots,x_n]$ be a ring of polynomials and
\[\begin{array}{cccc}
\gamma: & \K[x_0,\ldots,x_n] & \longrightarrow & A \\
 & f(x_0,\ldots,x_n) & \rightarrow & f(a_0,\ldots,a_n)
\end{array}\]
Let $H_\gamma$ be a finite subset of $\K[x_0,\ldots,x_n]$ which
generates $\Ker\ \gamma$. Let $B$ be a subalgebra of $A$ and
$\{b_1,\ldots,b_m\}$ a set of generators of $B$ given as polynomials
$B_i\in\K[x_0,\ldots,x_n]$ such that $\gamma(B_i)=b_i$. Let $c$ be
an element of $A$ given as a polynomial $C\in\K[x_0,\ldots,x_n]$
such that $\gamma(C)=c$.

\begin{theorem}\label{th-sweedler-tag-Kalgebras}
$ $
\begin{description}
    \item [(i)] If $A$ is an integral domain, given the field
extension $QF(A)/QF(B)$ it is possible to decide whether it is transcendental or algebraic and:
    \begin{itemize}
        \item if it is transcendental, its transcendence degree
can be computed;
        \item if it is algebraic, its degree can be computed.
    \end{itemize}

    \item [(ii)] It is possible to decide whether $c$ is integral
over $B$, and whether $c$ is algebraic over $QF(B)$ and:
    \begin{itemize}
        \item if it is algebraic, its minimum polynomial can be computed;
        \item in particular, we can determine whether $c\in QF(B)$
and in the affirmative case we can find an expression of $c$ in
terms of $b_i$.
    \end{itemize}
\end{description}
\end{theorem}

This theorem, which is stated for $\K$-algebras, has a simpler form
when our ambient field is rational.

\begin{corollary}\label{th-sweedler-tag}
We can compute transcendence and algebraic degrees of unirational
fields, decide whether an element is transcendental or algebraic
over a field, compute its minimum polynomial in the latter case, and
decide membership.
\end{corollary}

We illustrate this corollary with the following example:

\begin{example}\label{exam-sweedler}
Consider the rational functions $f_1, f_2$ in $\Q(x,y)$, where
\[f_1=-{y}^{2}x-{y}^{4}+2\,x+2\,{y}^{2}-1,\ f_2=4\,{y}^{4}-10\,{y}^{2}+5+3\,{y}^{2}x-6\,x.\]

We want to know if the field extension $\Q(x,y)/\Q(f_1,f_2)$ is algebraic or transcendental, and the corresponding
degree in each case. We compute a Gr\"obner basis $G$ of the ideal $I=(t_1-f_1,t_2-f_2)\subset\Q[x,y,t_1,t_2]$ with
respect to a tag monomial ordering $\{x,y\}>\{t_1,t_2\}$:
\[\begin{array}{rcl}
G & = & \{-3\,t_1+{y}^{4}-4\,{y}^{2}+2-{\it t_2},\\
 & & 3\,xt_1+x{t_2}+2\,x+4\,{y}^{2}{t_1}+{y}^{2}{t_2}+3\,{y}^{2}-2\,{t_1}-2,\\
 & & {y}^{2}x-2\,x+2\,{y}^{2}+4\,{t_1}+{t_2}-1\}.
\end{array}\]
so the transcendence degree is 2, because there is no polynomial involving only $t_1,t_2$.

On the other hand, the extension is algebraic of degree $4=4\times
1$. The polynomial $-3\,t_1+{y}^{4}-4\,{y}^{2}+2-{\it t_2}$ in $G$  indicates that $y$ is algebraic over
$\Q(f_1,f_2)$ and its minimum polynomial $z^4+z^2-3f_1-f_2+2$
has degree $4$.

Alternatively, a different Gr\"obner basis computed with respect to lex ordering with $y>x>t_1>t_2$ is
\begin{footnotesize}
\[\begin{array}{l}
\hspace{-5ex}\{12\,x{t_1}-16\,{t_1}^{2}-8\,{t_1}\,{t_2}-12\,{t_1}+3\,{x}^{2}{t_1}+{x}^{2}{t_2}+2\,{x}^{2}+8\,x+4\,x{t_2}-{t_2}^{2}-2\,{t_2}-1,\\
\hspace{-2ex} \ \ 3\,x{\it t_1}+x{\it t_2}+2\,x+4\,{y}^{2}{\it
t_1}+{y}^{2}{\it t_2}+3\,{y}^{2}-2\,{\it t_1}-2,\ -3\,{\it
t_1}+{y}^{4}-4\,{y}^{2}+2-{\it t_2}, \\
\hspace{-2ex} \ \ {y}^{2}x-2\,x+2\,{y}^{2}+4\,{ \it t_1}+{\it t_2}-1,\ -3\,{t_1}+{y}^{4}-4\,{y}^{2}+2-{t_2}\}\\
\end{array}\]
\end{footnotesize}
so $x$ is algebraic over $\Q(f_1,f_2)$ and its minimum polynomial has  degree 2.
\end{example}

The computations described in these theorems require Gr\"obner bases computation with respect tag orderings, thus the
computing time is (double) exponential in the number of variables and polynomial in the degree of $f_1,\dots,f_m$.

\subsection{Computation of separating bases}\label{subsect-steinwandt}

The results that we describe now will allow us to compute a separable basis and the transcendence degree of a
separable extension without computing Gr\"obner bases, greatly increasing the efficiency of our computations. See
\citep{Wei46} and \citep{Ste00} for more details about these techniques.

Let $\F=\K(g_1,\ldots,g_m)$ be a unirational field, $\K \subset \F \subset
\K(\x)$. First introduce a classical definition that will be very useful for
our purpose, see \citep{Wei46}.

\begin{definition}\label{def-ideal-rel}
Given a field extension $\K(\x)/\F$, we construct the ring homomorphism
$ \phi_\F:  \F[\y]  \longrightarrow  \K(\x)$ defined as $\phi_\F(y_i)= x_i$,
where $\y=(y_1,\ldots,y_n)$. Its kernel, which we will denote as
$\mathcal{B}_{\K(\x)/\F}$, is called the \emph{ideal of relations} of
the extension $\K(\x)/\F$.
\end{definition}

The paper \citep{MS99} presents a method to find explicit generators of the ideal by means of Gr\"obner bases
techniques. Because of this, the following theorem is fundamental, as it allows to express a related ideal (namely,
the extension of our ideal in a certain ring) in a very simple way.

We denote by $\F[\y]_{\mathcal{B}_{\K(\x)/\F}}$ the localization ring of $\F[\y]$ at the prime ideal
$\mathcal{B}_{\K(\x)/\F}$. Let $\mathcal{B}^e_{\K(\x)/\F}$ be the extended ideal of $\mathcal{B}_{\K(\x)/\F}$ in the
local ring $\F[\y]_{\mathcal{B}_{\K(\x)/\F}}$, \citep{AMc69}.

\begin{proposition} With the above notation, we have
\[\mathcal{B}^e_{\K(\x)/\F}=\langle g_1(\y)-g_1(\x),\ldots,g_m(\y)-g_m(\x)\rangle .\]
\end{proposition}

This result can be combined with the next Theorem to provide a
relatively fast way to compute transcendence degrees of separable
extensions.

\begin{theorem}\label{weil-t2}
Let $C=\{p_l=g_l(\y)-g_l(\x), \quad l=1,\ldots, m \}$ and $t=\td(\K(\x)/\F)$. Then
\[\rank\left(\frac{\partial p_i}{\partial y_j}(\x)\right)_{p_i\in C,j=1,\ldots,n}\leq n-t\]
and they are equal if and only if $\K(\x)/\F$ is separable.
\end{theorem}

\begin{corollary}\label{weil-t3}
With the notations of the previous theorem, if $I\subset C$ and
$J\subset\{1,\ldots,n\}$ are such that $\sharp I=\sharp J=n-t$ and
\[\det\left(\frac{\partial p_i}{\partial y_j}(\x)\right)_{p_i\in I,j\in J}\neq 0,\]
then the set $\{x_i:\ i\not\in J\}$ is a transcendence basis of
$\K(\x)/\F$.
\end{corollary}

We illustrate this with the following example.

\begin{example}\label{ej-steinwandt1}
Let
\[h_1=\frac{x_1+x_2-2\,x_3}{1+x_3x_2},\ h_2=\frac{x_1x_2-x_3}{x_1}\in \Q(x_1,x_2,x_3).\]
We construct the field $\Q(g_1,g_2,g_3,g_4)$ where
\[\begin{array}{lcl}
g_1 & = & \displaystyle\frac{x_1^2x_2+x_1x_2^2-3\,x_1x_2x_3-x_3x_1-x_3x_2+2\,x_3^2-x_1}{x_1^2+x_1x_2-2\,x_3x_1} = h_2-\displaystyle\frac{1}{h_1}, \\
g_2 & = & \displaystyle\frac{x_1^2x_2+x_1x_2^2-2\,x_1x_2x_3-x_3x_1-x_3x_2+2\,x_3^2}{x_1+x_1x_2x_3} = h_1h_2, \\
g_3 & = & \displaystyle\frac{x_1^2-x_1x_2-2\,x_3x_1+2\,x_3-2\,x_3x_2^2x_1+2\,x_3^2x_1}{x_1x_2-x_3+x_3x_2^2x_1-x_3^2x_2} =
 \displaystyle\frac{h_1}{h_2}-2, \\
g_4 & = & \displaystyle\frac{-x_1x_2+x_3-x_3x_2^2x_1+x_3^2x_2}{-x_1^2+2\,x_3x_1-x_3+x_3x_2^2x_1-x_3^2x_2} = \displaystyle\frac{h_1}{h_1-h_2} \\
\end{array}\]

It is clear that it has transcendence degree 2 over $\Q$. We have
\[C= \{ p_l=g_l(y_1,y_2,y_3)-g_l(x_1,x_2,x_3): l=1,\ldots,4 \} .\]

We construct the matrix $A=(a_{i,j})$, for $i=1,\ldots,3$ and  $j=1,\ldots,4$, where
\[a_{i,j}= \frac{\partial p_i}{\partial y_j}(x_1,x_2,x_3).\]

If we put it in triangular form we obtain:
\[\left(\begin{array}{ccc}
1 & 0 & -\displaystyle\frac{\left(x_1x_2^2+x_1^2x_2-2\,x_3^2+
x_3x_1+2\,x_1-1\right)x_1}{x_3x_2x_1^2+x_1^2+x_3^2x_1-x_3-2\,x_3^3}\\
\medskip 0 & 1 & \displaystyle\frac{x_3x_2^2-x_1+2\,x_3}{x_3x_2x_1^2+x_1^2+x_3^2x_1
-x_3-2\,x_3^3}\\
\medskip 0 & 0 & 0\\
\medskip 0 & 0 & 0
\end{array}\right)\]

The rank of the matrix is 2 as we expected. On the other hand, $x_3$
(the generator of the total field corresponding to the last column)
is a transcendence basis of $\Q(x_1,x_2,x_3)$ over $\Q(g_1,g_2,g_3,g_4)$.
\end{example}

\subsubsection{Jacobian matrix and uni-multivariate decomposition}

As an application of the results in this subsection, we will recover
the relation between the Jacobian matrix of a polynomial, see
\cite{Sha77}, and uni-multivariate decomposition, see \cite{GRS02}.

\begin{definition}
Given a list of polynomials $\Phi=(p_1,\ldots,p_n)$, where
$p_i\in\K[\x]$, we denote by $J(\Phi)$ the Jacobian matrix they
define, that is,
\[\begin{array}{cccc}
J(\Phi) & = & \left(
\begin{array}{cccc}
\displaystyle\frac{\partial p_1}{\partial x_1} & \displaystyle\frac{\partial p_1}{\partial x_2} &\ldots &\displaystyle\frac{\partial p_1}{\partial x_n}  \\
. & .& \ldots & . \\
. & .& \ldots & . \\
\displaystyle\frac{\partial p_n}{\partial x_1} & \displaystyle\frac{\partial p_n}{\partial x_2} &\ldots & \displaystyle\frac{\partial p_n}{\partial x_n}  \\
\end{array}
\right)
\end{array}\]
\end{definition}

Let $r=\td(\K(\x)/\K(p_1,\ldots,p_n))$. Assume that not every $p_i$
is constant, then $0\leq r\leq n-1$.

We will prove the following result:

\begin{theorem}\label{gamboa}
These statements are equivalent:
\begin{description}
    \item [(i)] There exist $f\in\K[\x]$, $q_i\in\K[t]$ such that
$p_i=q_i(f),\ i=1,\ldots,n$.
    \item [(ii)] The rank of the matrix $J(\Phi)$ is $n-1$.
\end{description}
\end{theorem}

First, we will translate this into a question about fields.

\begin{lemma}
The statement (i) in Theorem \ref{gamboa} is equivalent to $r=1$.
\end{lemma}

\begin{proof}
If (i) is true, then $\K[p_1,\ldots,p_n]\subset\K[f]$ and
$K(p_1,\ldots,p_n)\subset\K(f)$ so $\td(\K(p_1,\ldots,p_n)/\K)=1$.

Conversely, if $r=1$, by the Extended L\"uroth's Theorem we have
that\linebreak $\K(p_1,\ldots,p_n)=\K(f)$; as the field contains
some non-constant polynomial, by the same theorem we can assume
$f\in\K[\x]$. If suffices to prove for each $i$ that $p_i=q_i(f)$,
${q_i}_D\in\K^*$.

If $\gcd({q_i}_N,{q_i}_D)=1$, then for some $\alpha_i(t),\beta_i(t)\in\K[t]$ we have
\[1={q_i}_N(t)\alpha_i(t)+{q_i}_D(t)\beta_i(t)\ \Rightarrow\ 1={q_i}_N(f)\alpha_i(f)+{q_i}_D(f)\beta_i(f)\ \Rightarrow\]
\[\Rightarrow\ \gcd({q_i}_N(f),{q_i}_D(f))=1\ \Rightarrow\ {q_i}_D\in\K^*\ \Rightarrow\ q_i\in\K[t].\]
\end{proof}

Now consider the ideal of relations of $\K(\x)/\K(p_1,\ldots,p_n)$,
\[\mathcal{B}_{\K(\x)/\K(p_1,\ldots,p_n)}=\{h(\y)\in\K(p_1,\ldots,p_n)[\y]:\ h(\x)=0\}\]
where $\y=(y_1,\ldots,y_n)$ and $y_i$ are algebraically independent from $x_i$. Then we
have:

\begin{lemma}
Let $\overline{p}_i=p_i(\y)-p_i$. Then
\[\mathcal{B}_{\K(\x)/\K(p_1,\ldots,p_n)}=\langle \overline{p}_1,\ldots,\overline{p}_n \rangle.\]
\end{lemma}

\begin{proof}
``$\supset$'' is trivial. Conversely, given $h \in \mathcal{B}_{\K(\x)/\K(p_1,\ldots,p_n)}$ we can assume
$h\in\K[p_1,\ldots,p_n][\y]$. We write $h=\sum_{\alpha}h_{\alpha}(\x)\y^{\alpha}$,  where $h_{\alpha}(\x) \in
\K[p_1(\x),\ldots,p_n(\x)]$. Then $h(\x,\y)-\sum_{\alpha}(h_{\alpha}(\x)-h_{\alpha}(\y))\y^{\alpha}=h(\y,\y)$. Since
$h(\x,\x)=0$ we also have $h(\y,\y)=0$. We may write $h_{\alpha}(\x)=g_{\alpha}(p_1(\x),\ldots,p_n(\x))$ and do so for
$h_{\alpha}(\y)$ to get $g_{\alpha}(p_1(\y),\ldots,p_n(\y))$. It is then clear that
$g_{\alpha}(p_1(\y),\ldots,p_n(\y))-g_{\alpha}(p_1(\x),\ldots,p_n(\x))$ belongs the required ideal.
\end{proof}

%%%DAVID Puesto todo en una linea
Because of Theorem \ref{weil-t2}, if the extension $\K(\x)/\K(p_1,\ldots,p_n)$ is separable then
\[\rank \left(\begin{array}{cccc}
\displaystyle\frac{\partial \overline{p}_1}{\partial y_1}(\x) & \displaystyle\frac{\partial \overline{p}_1}{\partial y_2}(\x) & \ldots & \displaystyle\frac{\partial \overline{p}_1}{\partial y_n}(\x)  \\
. & . & \ldots & . \\
. & . & \ldots & . \\
\displaystyle\frac{\partial \overline{p}_n}{\partial y_1}(\x) & \displaystyle\frac{\partial \overline{p}_n}{\partial y_2}(\x) & \ldots & \displaystyle\frac{\partial \overline{p}_n}{\partial y_n}(\x)  \\
\end{array}\right)=n-r.\]

It is clear that the previous matrix is $J(\Phi)$ so the theorem we intend to prove is true if the
extension is separable. Besides, we cannot omit the hypothesis of separability, as the next
example shows.

\begin{example}
Let $\K=\F_p$, $p=x,q=y^p\in\F_p[x,y]$. Then
\[J(p,q)=\left(\begin{array}{cc}
  1 & 0 \\
  0 & 0 \\
\end{array}\right)\]
but $\td(\F_p(x,y^p)/\F_p)=2$.
\end{example}

Lastly, we have that $p_i=q_i(f)$, $i=1,\ldots,n$ if and only if
$\gcd(\overline{p}_i,\overline{q}_i)\neq 1$ for each $i$. Also, in
this case $\gcd(\overline{p}_i,\overline{q}_i)=\overline{f}$ where
$\overline{f}=f(u_1,\ldots,u_n)-f(\x)$ and
$\K[p_1,\ldots,p_n]=\K[f]$.

\section{The case of transcendence degree $n$}\label{sect-gradon}

Now we will study the case in which the extension $\K(\x)/\K(f_1,\ldots,f_m)$ is algebraic.

The problem of computing intermediate subfields in finite algebraic
extensions over the rational number field has been studied by
several authors, we can mention the paper \cite{LM85} and more
recently \cite{KP97}. Our approach it is a modification and adaptation
of \cite{LM85}'s techniques  and it is based on some general ideas
of Rubio's Ph. D. Thesis, \cite{Rub01}.

Corollary \ref{th-sweedler-tag} and Primitive Element Theorem allow us to rewrite the involved fields in the following
way:

\begin{itemize}
    \item There exist rational functions $\hat\alpha_1,\ldots,\hat\alpha_n$
such that $\K(\hat\alpha_1,\ldots,\hat\alpha_n)/\K$ is a purely
transcendental extension, with
\[\K(\hat\alpha_1,\dots,\hat\alpha_n)\subset\K(f_1,\dots,f_m)\subset\K(x_1,\dots,x_n).\]

    \item There exist $\hat\alpha_{n+1},f$ algebraic over
$\K(\hat\alpha_1,\ldots,\hat\alpha_n)$ such that
\[\begin{array}{ll}
\K(f_1,\dots,f_m)= & \K(\hat\alpha_1,\ldots,\hat\alpha_n,\hat\alpha_{n+1}), \\
\K(x_1,\dots,x_n)= & \K(\hat\alpha_1,\dots,\hat\alpha_n,f).
\end{array}\]
\end{itemize}

Also, for any intermediate field in the extension there exists $h$
algebraic over $\K(\hat\alpha_1,\ldots,\hat\alpha_n)$ such that
\[\F=\K(\hat\alpha_1,\ldots,\hat\alpha_n,h).\]

The structure of the lattice of intermediate fields in the extension\linebreak $\K(\x)/\K(f_1,\dots,f_m)$ suggests the
following diagram: let
\[\begin{array}{cccc}
\Phi: & \K(\hat\alpha_1,\ldots,\hat\alpha_n) & \longrightarrow & \E=\K(t_1,\dots,t_n) \\
 & \hat\alpha_i & \longmapsto & t_i
\end{array}\]
where $t_1,\dots,t_n$ are new free variables. $\Phi$ is an isomorphism that can be extended to $\K(\x)$ by means of an
isomorphism $\hat{\Phi}$:
\[\begin{array}{cccc}
\hat{\Phi}: & \K(\x) & \longrightarrow & \E[\alpha] \\
 & \hat\alpha_i & \longmapsto & t_i \\
 & f & \longmapsto & \alpha
\end{array}\]

We have that $\hat{\Phi}(\K(\x))$ is algebraic over $\E$. By the Primitive Element Theorem, we can write
$\hat{\Phi}(\K(\x))=\E[\alpha]$, where $\alpha$ is algebraic over $\E$. $\hat{\Phi}$ is an isomorphism that extends
$\Phi$.

On the other hand, $f$ is algebraic over $\K(\hat\alpha_1,\ldots,\hat\alpha_n)$. Then there exists its minimum
polynomial $p_f(\hat\alpha_1,\ldots,\hat\alpha_n,z)$ and it can be computed with Corollary \ref{th-sweedler-tag}. As
$\hat {\Phi}$ is an isomorphism, $p_f(t_1,\dots,t_n,z)$ is the minimum polynomial of $\alpha$ over $\E$ and
$\E[\alpha]=\E[z]/(p_f)$.

Once we have the isomorphism $\hat{\Phi}$, it can be restricted to $\K(f_1,\ldots,f_m)$ or any intermediate field $\F$
of $\K(\x)/\K(f_1,\dots,f_m)$. Analogously we have
\[\begin{array}{cccc}
\hat{\Phi}| \F: & \F & \longrightarrow & \E[\gamma] \\
 & \hat\alpha_i & \longmapsto & t_i \\
 & h & \longmapsto & \gamma
\end{array}\]
where $\gamma$ is algebraic over $\E$ with minimum polynomial
$p_h(t_1,\dots,t_n,z)$.

Conversely, given a field $\E[\gamma]$ such that
$\E[\beta]\subset\E[\gamma]\subset\E[\alpha]$, the inverse of
$\hat{\Phi}$ gives the intermediate field
$\F=\K(\hat\alpha_1,\ldots,\hat\alpha_n)(\hat{\Phi}^{-1}(\gamma))$.

The resulting diagram is
\begin{diagram}\label{diagram}
$ $
\[\begin{array}{cccc}
\K(x_1,\dots,x_n) & \longleftrightarrow & \E[\alpha]=\E[z]/(p_f) \\
\uparrow & & \uparrow \\
\F & \longleftrightarrow & \E[\gamma]=\E[z]/(p_h) \\
\uparrow & & \uparrow \\
\K(f_1,\dots,f_m) & \longleftrightarrow & \E[\beta]=\E[z]/(p_{\hat\alpha_{n+1}}) \\
\uparrow & & \uparrow \\
\K(\hat\alpha_1,\dots,\hat\alpha_n)& \longleftrightarrow & \E
\end{array}\]
\end{diagram}

This diagram is interesting because we can decide computationally
the inclusion of these fields.

\begin{theorem}
Let $\E[\alpha]/\E$ be an algebraic extension and
$\E[\beta],\E[\gamma]\subset\E[\alpha]$ intermediate fields. Then we
can decide if $\E[\beta]\subset\E[\gamma]$.
\end{theorem}

\begin{proof}
A subfield $\E[\beta]$ of $\E[\alpha]$ is determined by means of the minimum polynomial of $\beta$ over $\E$,
$p_\beta$, and by a polynomial $f\in\E[x]$ such that $\beta=f(\alpha)$. If $\E[\beta]\subset\E[\gamma]$, then
$\beta=p(\gamma)$ where $\deg\ p<\deg\ p_\gamma$, that is, $\beta=a_{l-1}\gamma^{l-1}+\cdots+a_0$. On the other hand,
$\beta,\gamma\in\E[\alpha]$, so deciding if $\E[\beta]\subset\E[\gamma]$ can be done by solving a system of linear
equations with $\deg\ p_\alpha$ equations (as $\{1,\alpha,\ldots,\alpha^{\deg\ p_\alpha-1}\}$ is a basis of the
$\E$-vector space $\E[\alpha]$), and $\deg\ p_\gamma$ variables $a_{l-1},\dots,a_0$.
\end{proof}

In \citep{LV93} there is another method to decide field inclusion
using resolvents when $\E=\Q$.

As a consequence we have that the problem is solved for fields with
characteristic zero if we can find all intermediate fields of the
algebraic extension $\E[\alpha]/\E$. Now we will study how to find
those fields.

We will denote by $\LL=\E[\alpha_1,\dots,\alpha_m]$ the splitting field of $\E[\alpha]$, being
$\alpha=\alpha_1$. Due to Galois Theory we know that there is a bijection between the lattice of
intermediate fields of $\E\subset\LL$ and the subgroups of the Galois group of $\E\subset\LL$,
which we will denote as $G$. If we define $G_\alpha=\{\sigma\in G:\ \sigma(\alpha)=\alpha\}$,
there is also a bijection between the subgroups of $G_\alpha\subset G$ and certain roots of the
minimum polynomial $p_\alpha$ of $\alpha$. These correspondences are the key to the method that we
present to find intermediate fields of simple algebraic extensions. First, we present an adapted
version of the classical fundamental theorem of Galois theory.

\begin{theorem}
There exists a bijection between the set of intermediate fields of
$\E\subset\E[\alpha]$ and the set of subgroups of $G$ that contain
$G_\alpha$.
\end{theorem}

So, we can work with the Galois group of the extension, for which we
will use the so called decomposition blocks, that we introduce now,
see \citep{Wie64}.

\begin{definition}
Let $f\in\K[x]$ be an irreducible polynomial, $G$ the Galois group
of $f$ over $\K$ and $\Omega=\{\alpha=\alpha_1,\ldots,\alpha_m\}$
the set of roots of $f$.

A subset $\psi\subset\Omega$ is a \emph{decomposition block} if for
each $\sigma\in G$ we have either $\sigma(\psi)\cap\psi=\emptyset$
or $\sigma(\psi)=\psi$.

The blocks $\{\alpha_i\}$ and $\Omega$ are called \emph{trivial
blocks}.

The set of blocks that are conjugate to $\psi$, that is
$\psi,\sigma_2(\psi),\ldots,\sigma_r(\psi)$, are a \emph{block
system}.

If $|\psi|=s$ we say that the block $\psi$ is a $r\times
s$-\emph{decomposition block}, where $(m=rs)$.
\end{definition}

The next theorem gives a bijection between the intermediate groups
of $G_\alpha\subset G$ and the decomposition blocks that contain
$\alpha$. The proof is an adaptation of the one in \citep{Wie64}.

\begin{theorem}
There exists a bijection between the set of intermediate groups of
$G_\alpha\subset G$ and the set of decomposition blocks that contain
$\alpha$. Besides, the correspondence respects inclusions.
\end{theorem}

\begin{proof}
We define the following bijection:
\[\begin{array}{crc}
\{H:\ G_\alpha\subset H\subset G\} & \longrightarrow & \{\psi:\ \alpha\in\psi\} \\
H & \longmapsto & \psi_H=\{\sigma(\alpha):\ \sigma\in H\}
\end{array}\]

In order to see that it is well defined we must prove that $\psi_H$
is a decomposition block. Let $\sigma\in G$ and assume that
$\beta\in\sigma(\psi_H)\cap\psi_H$. By definition there exist
$\tau_1,\tau_2\in H$ such that
$\beta=\tau_1(\alpha)=\sigma(\tau_2(\alpha))$, this implies that
$\tau_1^{-1}\sigma\tau_2\in G_\alpha\subset H$. In this way we have
that $\sigma\in H$ and thus $\sigma(\psi_H)=\psi_H$. Also,
$\alpha\in\psi_H$.

Now let $H_1,H_2$ be subgroups of $G_\alpha\subset G$ such that
$\psi_{H_1}=\psi_{H_2}$. If $\sigma\in H_1$, there exists $\tau\in
H_2$ with $\sigma(\alpha)=\tau(\alpha)$. Then $\tau^{-1}\sigma\in
G_\alpha\subset H_2$ and so $\sigma\in H_2$.

Let $\psi$ be a decomposition block with $\alpha\in\psi$. The
inverse image of $\psi$ is the subgroup $H=\{\sigma\in G:\
\sigma(\psi)=\psi\}$. Indeed, $H$ is a subgroup and $G_\alpha\subset
H$. We will see that $\psi=\psi_H$:

Let $\beta\in\psi$. As $G$ is transitive there exists $\sigma\in G$
such that $\beta=\sigma(\alpha)$. On the other hand,
$\alpha,\beta\in\psi$, so $\sigma\in H$ and $\beta\in\psi_H$.
Conversely, if $\beta\in\psi_H$, there exists $\sigma\in H$ such
that $\beta=\sigma(\alpha)$, and as $\sigma(\psi)=\psi$ we have
$\beta\in\psi$.

It is trivial that this bijection respects inclusions.
\end{proof}

The correspondences described so far allow us to construct the
following diagram:

\[\begin{array}{ccccc}
\LL & \longleftrightarrow & \{id\} & & \\
\uparrow & & \downarrow & & \\
\E[\alpha] & \longleftrightarrow & G_\alpha & \longleftrightarrow & \{\alpha\} \\
\uparrow & & \downarrow & & \downarrow \\
\F & \longleftrightarrow & H & \longleftrightarrow & \{\alpha_{i_1},\ldots,\alpha_{i_j}\} \\
\uparrow & & \downarrow & & \downarrow \\
\E & \longleftrightarrow & G & \longleftrightarrow &
\{\alpha_1,\dots,\alpha_m\}
\end{array}\]

It is important to highlight that, given a decomposition block
$\psi$, we can directly compute the corresponding field $\F_\psi$
without computing the corresponding group.

\begin{theorem}\label{bloque-a-field}
Let $\psi=\{\alpha_{i_1},\ldots,\alpha_{i_k}\}$ be a decomposition
block, then the corresponding field in the previous diagram is
$\E[\beta_1,\ldots,\beta_k]$ where each $\beta_j$ is the $j$-th
elementary symmetric function in $\alpha_{i_1},\ldots,\alpha_{i_k}$.
\end{theorem}

\begin{proof}
Let
$h(z)=\prod_{j=1}^k(z-\alpha_{i_j})=z^k+a_{k-1}z^{k-1}+\cdots+a_0$,
then
\[\E[\beta_1,\ldots,\beta_k]=\E[a_{k-1},\ldots,a_0].\]

We will see that $\E[a_{k-1},\ldots,a_0]=\E_\psi$:

Let $\sigma\in G_\psi$, then $\sigma(h)=h$ and $\sigma(a_i)=a_i$ for
every $i$. That is, $\E[a_{k-1},\dots,a_0]\subset\E_\psi$.

Now let $\sigma\in G_{\E[a_{r-1},\ldots,a_0]}$, then
$\sigma(a_i)=a_i$ for each $i$. Therefore $\sigma(h)=h$ and
$\sigma(\psi)=\psi$, and $\E_\psi\subset\E[a_{k-1},\ldots,a_0]$.
\end{proof}

Next, we will show the results that will support the algorithm that
solves our problem.

\begin{lemma}
Let $q(z,\alpha)$ be an irreducible factor of $p_\alpha$, the
minimum polynomial of $\alpha$, and $\psi$ a decomposition block
that contains $\alpha$. If a root of $q(z,\alpha)$ is in $\psi$,
then all the roots of $q(z,\alpha)$ are in $\psi$.
\end{lemma}

Depending on the factorization of the minimum polynomial we have
different methods to compute the decomposition blocks. We will
assume that
\[p_\alpha(z)=(z-\alpha)p_2(z,\alpha)\cdots p_l(z,\alpha).\]

Among the next results, the first one is interesting in itself,
because we can easily compute the intermediate fields when the
extension is normal.

\begin{theorem}
If $\E[\alpha]/\E$ is normal, we can compute all the intermediate fields; and one of them on
polynomial time if the algebraic degree of the extension is not prime.
\end{theorem}

\begin{proof}
Assume that
$p_\alpha(z)=(z-\alpha)(z-p_2(\alpha))\cdots(z-p_l(\alpha))$. Then
the Galois group of $p_\alpha$ over $\E$ is
$G=\{\sigma_i:\alpha\mapsto p_i(\alpha),\ i=1,\dots,l\}$. Each
subgroup $H$ of $G$ corresponds with a subfield
$\F=\E[a_0,\dots,a_{l-1}]$ with
$x^l+a_{l-1}x^{l-1}+\cdots+a_0=\prod\limits_{\sigma\in
H}(z-\sigma(\alpha))$.

Also, a group $G$ has non-trivial subgroups if and only if
$|G|=l=[\E[\alpha]:\E]$ is composite.
\end{proof}

\begin{theorem}
If the extension $\E[\alpha]/\E$ is not normal and $p_\alpha$ has
more than one root in $\E[\alpha]$, there exists a field $\F$ such
that $\E\varsubsetneq\F\varsubsetneq\E[\alpha]$.
\end{theorem}

\begin{proof}
If
\[p_\alpha(z)=(z-\alpha)(z-p_2(\alpha))\cdots(z-p_{l'}(\alpha))\cdot p_{l'+1}(z,\alpha)\cdots p_l(z,\alpha)\]
is the complete factorization of $p_\alpha$, then
$H=\{\sigma_i:\alpha\mapsto p_i(\alpha),\ i=1,\dots,l'\}$ is a
subgroup of $G$. Indeed, let $\sigma_i,\sigma_j\in H$, then
\[\begin{array}{rcl}
p_\alpha(z) & = & \sigma_j(p_\alpha(z))  \\
 & = & \sigma_j((z-\alpha)(z-p_2(\alpha))\cdots(z-p_{l'}(\alpha))p_{l'+1}(z,\alpha)\cdots p_l(z,\alpha)) \\
 & = & (z-p_j(\alpha))(z-p_2(p_j(\alpha)))\cdots(z-p_{l'}(p_j(\alpha)))\cdots \\
 & & \cdots p_{l'+1}(z,p_j(\alpha))\cdots p_l(z,p_j(\alpha))
\end{array}\]
is another factorization of $p_\alpha$ in $\E[\alpha]$. Then there
exists $k\in\{1,\dots,l'\}$ such that
$\sigma_i\sigma_j(\alpha)=p_i(p_j(\alpha))=p_k(\alpha)=\sigma_k(\alpha)$.
Therefore, $\langle H\cup G_\alpha \rangle$ is a subgroup of
$G_\alpha\subset G$; and it is non-trivial since $G$ is transitive
over the roots of $p_\alpha$ and $\langle H\cup G_\alpha \rangle$ is
not.

Because of this, $\E[a_0,\dots,a_{l'-1}]$ is an intermediate field
of $\E\subset\E[\alpha]$, being
\[x^{l'}+a_{l'-1}x^{l'-1}+\cdots+a_0=\prod_{i=1}^{l'}(z-p_i(\alpha)).\]
\end{proof}

The remaining case is that in which $p_\alpha$ has exactly one
linear factor. In this case, one must combine the factors of
$p_\alpha$ to check which of those divisors provide an intermediate
field. In the worst case, we must check an exponential number of
factors; but in other cases, we can find subfields even if we don't
have the complete factorization of $p_\alpha$.

%%%DAVID Reescrito un poco la ultima frase
As it is made clear before, we need to factorize polynomials whose coefficients are in some
algebraic extension of the field we work in. Next, we will give the details of a method to compute
such a factorization. We will show that the algorithm in \citep{Tra76} that factors polynomials is
in random polynomial time if the base field is a rational function field over $\K$ and there is a
polynomial time algorithm to factorize univariate polynomials over $\K$.

We will adapt the algorithm to fields $\E$, $\F$ where $\F/\E$ is a
finite algebraic separable extension. We will also present some
slightly shorter proofs of some results. The idea is similar to the
one presented in \citep{Wae64}, but more efficient from the
computational point of view. It is based on the fact that the
polynomial $f(x-c\alpha)\in\E[\alpha]$ and its norm have essentially
the same factorization unless the norm is not square free. Trager's
reduction is used in \citep{Lan85} to provide an algorithm in
polynomial time to factorize polynomials in algebraic number fields,
using the known univariate factorization algorithm over the
rationals in \citep{LLL82}.

The situation we are interested in is given by a field extension
\[\E\subset\E(\alpha_1,\dots,\alpha_m)=\F\]
that is algebraic and separable. This satisfies the hypothesis of the Primitive Element Theorem; a
constructive version for the case $\E=\Q$ is in \citep{YNT89}. The proof for an arbitrary
algebraic extension is similar. Other methods can be found in \citep{Loo83}.

In the following we will use these notations:

\begin{notation} $ $
\begin{itemize}
    \item $\F/\E$ is a finite separable algebraic extension.
    \item Due to the Primitive Element Theorem we can write
$\F=\E[\alpha]$.
    \item $p_\alpha$ is the minimum polynomial of $\alpha$
over $\E$.
    \item $\alpha_1,\dots,\alpha_l$ are the roots of $p_\alpha$
in $\overline{\E}$, the algebraic closure of $\E$.
    \item $G$ is the Galois group of $p_\alpha$ over $\E$.
\end{itemize}
\end{notation}

Let us remember the definition of the norm of a polynomial:

\begin{definition}
Let $f(\alpha,x)\in\F[x]$. We define the \emph{norm} of $f$ as
\[\Nor(f)=\prod_{i=1}^l\ f(\alpha_i,x).\]
\end{definition}

Using the known properties of the resultant, we have
 \[\Nor(f)=\Res_t(p_\alpha(t),f(t,x))\in\E[x].\]

The following is a classical result about the norm.

\begin{proposition}
Let $f(\alpha,x)\in\F[x]$ be an irreducible polynomial. Then
$\Nor(f)$ is a power of an irreducible polynomial over $\E$.
\end{proposition}

The key result in \citep{Tra76} is the following:

\begin{theorem}
Let $f(\alpha,x)\in\F[x]$ be an irreducible polynomial such that
$\Nor(f)$ is square free. If $\Nor(f)=h_1\cdots h_m$ is a complete
factorization in $\E[x]$, then
\[f=\gcd(h_1,f)\cdots\gcd(h_m,f)\]
is a complete factorization of $f$ in $\F[x]$.
\end{theorem}

With the results we have presented, we can factorize a polynomial
over $\F$ if $\E$ has an algorithm for univariate factorization,
except when the norm of the polynomial is not square free. To avoid
this we can apply a map $x\mapsto x-c\alpha$. Such a map always
exists because of a simple result due to Kronecker.

\begin{theorem}[Kronecker]\label{th-libre-cuad}
Let $f(\alpha,x)\in\F[x]$ be a square free polynomial with degree
$m$. Assume that $l=[\F:\E]$ and $\E$ has more than
$\displaystyle\frac{l(l-1)m(m-1)}{2}$ non-zero elements. Then there
exists $c\in\E$ such that $\Nor(f(\alpha,x-c\alpha))$ is square
free.
\end{theorem}

The combination of these results provides the following
factorization algorithm.

\begin{algo}\label{alg-factor-ext} $ $
\begin{description}
    \item \textsc{Input}: $f(\alpha,x)\in\E[\alpha][x]$.
    \item \textsc{Output}: a complete factorization
$f_1(\alpha,x),\ldots,f_m(\alpha,x)$ of $f(\alpha,x)$ in
$\E[\alpha][x]$.
\end{description}
\begin{description}
    \item \textsc{A}. Find $c\in\E$ such that $\Nor(f(\alpha,x-c\alpha))$
is square free.

    \item \textsc{B}. Factor $\Nor(f(\alpha,x-c\alpha))$
in $\E[x]$ to obtain a complete factorization
\[\Nor(f(\alpha,x-c\alpha))=h_1\cdots h_m.\]

    \item \textsc{C}. Compute $f_i=\gcd(f,h_i(x+c\alpha))$.
Return the $f_i$'s.
\end{description}
\end{algo}

\begin{analysis}
We will analyze the algorithm in our particular setting. We are
interested in $\E$ being a rational function field $\E=\K(\x)$ over
$\K$. Factorization over $\K(\x)[x]$ is equivalent to factorization
in the ring of polynomials $\K[x_1,\dots,x_n,x]$. On the other hand
it is known that every factorization algorithm in polynomial time in
$\K[x]$ provides one in random polynomial time in
$\K[x_1,\dots,x_n][x]$, using Hilbert's Irreducibility Theorem, see
\citep{GG99} and \citep{Zip93}.

Also, if the number of variables is zero ($n=0$), the previous
result by Kronecker requires that the field $\K$ has at least
$l^2m^2$ elements, where $m$ is the degree of the polynomial and
$l=[\F:\E]$. If $n>0$ there is always an adequate $\alpha\in\E[x]$,
as $\E$ is infinite.

Finally, step \textsc{C} requires the computation of several gcd's
in $\E[\alpha]$. This is also in polynomial time due to Euclides'
Algorithm, for more details of this part see \citep{LM85} for $\Q$.
\end{analysis}

Summarizing the results we have presented in this section, we have
the following algorithm to find intermediate unirational fields over
a given field, if the extension is separable and algebraic.

With the above notation:

%%%DAVID Cambiado el formato para que aparezca igual que los otros algoritmos
\begin{algo}\label{alg-algebraic} $ $
\begin{description}
    \item \textsc{Input}: An irreducible $f(t) \in \E[t]$, such that $f(\alpha)=0$ and  $p_\alpha(z) \in E[\alpha][z]$.
    \item \textsc{Output}: All $h(t) \in \E[t]$ such that $\E[h(\alpha)] \subset\E[\alpha]$.
\end{description}
\begin{description}
    \item \textsc{A}. Factorize $p_\alpha(z)$ in $E[\alpha]$.

    \item \textsc{B.1}. If $p_\alpha(z)$ has more than one linear factor:
\[p_\alpha(z)=(z-\alpha)(z-p_2(\alpha))\cdots(z-p_r(\alpha)) p_{r+1}(z,\alpha)\cdots
p_{r'}(z,\alpha).\]
\begin{description}
    \item --- Compute a minimal subgroup $G_{\psi}$ of $\langle\{\sigma_2:\alpha\mapsto p_i(\alpha)\}\rangle$.
    \item --- Consider $h(z)=\prod_{\sigma\in G_{\psi}}(z-\sigma(\alpha))=a_ux^u+\cdots+a_0$.
    \item --- Take $a_i$ such that $\E[a_i]$ is a proper subfield of $\E\subset\E[\alpha]$.
\end{description}

    \item \textsc{B.2}. If $p_\alpha(z)=(z-\alpha)p_2(z,\alpha)\cdots p_{r'}(z,\alpha)$, with $p_i$
    non-linear:
\begin{description}
    \item --- Consider a factor $P_2(z)=h(z, \alpha)(z-\alpha)$ of $p_\alpha(z)$,
\[P_2=(z-\alpha)h(z,\alpha)=a_ux^u+\cdots+a_0.\]
    \item --- If $\E[a_i]= \E[\alpha]$ for all $i$, then  take another factor.
\end{description}
\end{description}
\end{algo}

We illustrate this algorithm with the following example:

\begin{example}
Consider the rational functions $f_1, f_2$ in $\Q(x,y)$ in Example \ref{exam-sweedler}
\[f_1=-{y}^{2}x-{y}^{4}+2\,x+2\,{y}^{2}-1,\ f_2=4\,{y}^{4}-10\,{y}^{2}+5+3\,{y}^{2}x-6\,x.\]

Our goal is computing all intermediate fields in the extension $\Q(x,y)/\Q(f_1,f_2)$.

By Example \ref{exam-sweedler}, we know it is an algebraic extension of degree $4$. Moreover, $y$
is a primitive element and its minimum polynomial is
\[p_y(f_1,f_2,z)=z^4+z^2-3f_1-f_2+2.\]

Clearly, if $\alpha$ is a root of $p_y(t_1,t_2,z)$, then also
$-\alpha$ is a root, so we have a factorization
\[p_y(t_1,t_2,z)=(z-\alpha)(z+\alpha)(z^2+\alpha^2-4)\]
in $E[\alpha]=E[z]/(p_y)$.

Let $H=\{id,\alpha\to -\alpha\}$ and $h(z)=z^2-\alpha^2$, we obtain
the proper field $\E\subset\E[\alpha^2]\subset\E[\alpha]$
\[\Q(f_1,f_2)\varsubsetneq\Q(f_1,f_2,y^2).\]

To determine all intermediate fields, we need to factorize
$p_y(t_1,t_2,z)=(z-\alpha)(z+\alpha)(z^2+\alpha^2-4)$. In order to do this we will use Algorithm
\ref{alg-factor-ext}. As the polynomial $g(z,\alpha)=z^2+\alpha^2-4$ divides the polynomial
$p_y(t_1,t_2,z)$, we apply a transformation (see Theorem \ref{th-libre-cuad}), for example $z\to
z-3\alpha$. The next step is computing the norm of $g(z-3,\alpha)$.

\[\begin{array}{rcl}
N(g(z-3\alpha,\alpha)) & = & \Res_z(p_y(t_1,t_2,z),(z-3\alpha)^2+\alpha^2-4) \\
 & = & 4-4\,{\it t_2}+6\,{\it t_1}\,{\it t_2}+{{\it t_2}}^{2}-12\,{\it t_1}-1568\, {\alpha}^{2}+10784\,{\alpha}^{4} \\
 & & +9\,{{\it
t_1}}^{2}-1104\,{\it t1}\,{ \alpha}^{2}-816\,{t_1}\,{\alpha}^{4}-368\,{\it t_2}\,{\alpha}^{2} \\
 & & -272\,{\it t_2}\,{\alpha}^{4}-13312\,{\alpha}^{6}+4096\,{\alpha}^{8}.\\
\end{array}\]

As $N(g(z-3\alpha,\alpha))$ is irreducible, also $z^2+\alpha^2-4$ is
and we already have a complete factorization of the minimum
polynomial. Therefore, the extension is not normal and in order to
find more intermediate fields we only have to consider the divisor
$(z-\alpha)(z^2+\alpha^2-4)$; but it cannot provide a decomposition
block, as 3 does not divide the degree of the extension.

The lattice of fields is then
\[\Q(f_1,f_2)\varsubsetneq\Q(f_1,f_2,y^2)\varsubsetneq\Q(x,y).\]

Finally, we note that the intermediate field we found is rational,
in fact
\[\Q(f_1,f_2,y^2)= \Q(x-y,x+y^2)=\Q(x,y^2).\]
However, as we said, our algorithm always returns a number of
generators which is equal to the transcendence degree plus one (see
Theorem \ref{cota-gens}).
\end{example}

\subsection{Normality and monodromy group}

The computation of intermediate fields is even more interesting and simpler when the algebraic
extension $\K(\x)/\K(f_1,\ldots,f_m)$ is normal. In this case we have the known bijection between
subgroups of the Galois group and intermediate fields. We will now concentrate on the case $n=1$
and assume that $\car\ \K=0$. Our problem can be stated in the following way:

\begin{problem}
Given an irreducible polynomial $f\in\K[x]$, determine if the
extension $\K(\alpha)/\K$, where $\K(\alpha)=\K[x]/(f)$, is normal.
\end{problem}

We can use simple Galois techniques to decide this matter. Remember that we assume that the
extension $\K(x)/\K(f)$ is algebraic.

\begin{definition}
Let $f\in\K(x)$. We call \emph{monodromy group} of $f$ to the Galois group of the extension
$\K(x)/\K(f)$. That is, if we denote by F the splitting field of the extension $\K(x)/\K(f)$, then
the monodromy group of $f$ is the group of automorphisms of $\F$ that leave $\K(f)$ fixed.
\end{definition}

\begin{theorem}
The extension $\K(x)/\K(f)$ is normal if and only if
$G(f)=\{u\in\Aut_\K\,\K(x): f\circ u=f\}$ is equal to the monodromy
group of the extension.
\end{theorem}

\begin{proof}
the roots of the minimum polynomial are the images of one of them
through the elements of the Galois group; if it is equal to $G(f)$,
they are all in $\K(f)$.
\end{proof}

\begin{corollary}
The extension $\K(x)/\K(f)$ is normal if and only if $|G(f)|=\deg\
f$.
\end{corollary}

Also, the techniques for factorization in algebraic extensions that we discussed above provide
another method: we factorize the polynomial $f$ in $\K(\alpha)$, then the extension is normal if
and only if $f$ splits in this field.

\begin{remark}
If the extension is normal, factorization in extensions is actually
performed over the base field, which greatly improves the efficiency
of the algorithm.
\end{remark}

Finally, we can also try to decide normality simply by writing the
corresponding equations. In particular, the extension is normal if
and only if all the roots of $f$ are in $\K(\alpha)$. There is a
known bijection between the polynomials $p(x)\in\K[x]$ with $\deg\
p\leq\deg\ f$ and the elements of $\K(\alpha)$, namely the morphism
$x\rightarrow\alpha$ from $\K[x]$ to $\K(\alpha)$; therefore, each
$p$ represents a root of $f$ in $\K(\alpha)$ if and only if $f(p)=0$
in $\K(\alpha)$, that is , $f(x)$ divides $f(p(x))$ in $\K[x]$. This
is precisely the classic problem of ideal decomposition, see
\citep{CFM96}.

The previous relation can be expressed directly with equations: let
\[\begin{array}{ccl}
f & = & x^n+a_{n-1}x^{n-1}+\cdots+a_1x+a_0, \\
p & = & b_{n-1}x^{n-1}+\cdots+b_1x+b_0, \\
q & = & x^m+c_{m-1}x^{m-1}+\cdots+c_1x+c_0, \quad m=n(n-2).
\end{array}\]

Then from the expression $f(p)=f\cdot q$ we obtain a linear system
of equations in the variables $b_i$ and $c_j$. Note that we are only
interested in the existence and computation of values for the
variables $b_i$.

As indicated in the introduction, the particular case in which the
given field has transcendence degree one over $\K$ was solved in
\citep{GRS01}.

\section{The general case and its reduction to the algebraic case}\label{sect-casogeneral}

Our strategy for the resolution of the general problem comes down to reducing it to the case where
the given field has transcendence degree $n$ over $\K$, that is, the extension
$\K(\x)/\K(f_1,\ldots,f_m)$ is algebraic. To that end we will present two different methods, and
also the outline of another one.

\subsection{Relative algebraic closure}\label{subsect-clausura}

We will look for the minimum field that contains all the
intermediate algebraic fields over the given one. To this end we
adapt the method in \citep{BV93} and in the recent book
\citep{Vas98} to compute the closure of a ring monomorphism.

\begin{definition}
Let $R_1\subset R_2$ be a ring extension. We call \emph{integral
closure of $R_1$ relative to $R_2$} to the subring of $R_2$ formed
by the elements that are integral over $R_1$.
\end{definition}

%%%DAVID La primera frase estaba suelta mas abajo asi que no se definia F_0
In our case, we need to compute the algebraic closure $\F_0$ of the field extension
$\K(f_1,\ldots,f_m)\subset\K(\x)$. Our goal is to determine explicitly a finite set of generators
of the field $\F_0$, in particular as many as the transcendence degree plus one. The Theorem
\ref{cota-gens} proves that such a set exists.

There are several methods to compute the integral closure of an
integral domain in its field of fractions, see for example
\citep{Sei75} and the more recent \citep{GT97}. The idea in
\citep{BV93} is to compute the integral closure of a birational
morphism:

\begin{theorem}
Let $\D_1\subset\D_2$ be an extension of integral domains that are finitely generated over a
computable field $\K$ with the same field of fractions (that is, a \emph{birational morphism}).
Let $\overline\D_1$ be the integral closure of $\D_1$ in its field of fractions. Assume that
$\D_2$ is generated over $\D_1$ by fractions whose denominators are powers of some element $d$.
Let $r$ be such that $\overline\D_1d^{r+1}\cap\D_1\subset(d)$. Then the integral closure of $\D_1$
in $\D_2$ is
\[d^{-r}(d^r\D_2\cap d^r{\overline\D_1}\cap\D_1).\]
\end{theorem}

We are in the most general situation, that is, $\D_1\subset\D_2$ is
an extension of integral domains that are finitely generated over a
computable field $\K$. We will follow these steps,see \citep{Vas98}:

\begin{description}
    \item 1. We write $\D_2=\D_1[b_1,\ldots,d_r].$

    \item 2. Let $t$ be a new variable and
$\D=\D_1[t,b_1,\ldots,d_r]\subset\D_2[t]$. It is a birational monomorphism.

    \item 3. We compute the integral closure $\overline\D$ of the extension
$\D\subset\D_2[t]$ according to the previous theorem.

    \item 4. Then the integral closure of the extension $\D_1\subset\D_2$ is
\[\overline\D\cap\D_2.\]
\end{description}

First we reduce the problem to integral closures of the corresponding integral domains:

\begin{theorem}
Let $\D_1\subset \D_2$ be two integral domains. Let $\D$ be the
integral closure of $\D_1$ with respect to $\D_2$. Let $\K_1$ and
$\K_2$ be the fields of fractions of $\D_1$ and $\D_2$
respectively and $\K$ the algebraic closure of $\K_1$ with respect
to $\K_2$. Then $\K$ is the field of fractions of $\D$.
\end{theorem}

\begin{proof}
Let $S=\D_1^*$ be the closed multiplicative system of non-zero elements in the integral domain
$\D_1$. Then $S^{-1}\D$ is, see \citep{AMc69}, the integral closure of
\[S^{-1}\D_1\subset S^{-1}\D_2.\]

As $\K_1=S^{-1}\D_1\subset S^{-1}\D$ is integral and $S^{-1}\D_1$ is a field, then $S^{-1}\D$ is a
field. Indeed, let $\alpha$ be an integral element over $\K_1$; dividing the equation by a power
of $\alpha$, we can write $\alpha^{-1}$ as an element of $S^{-1}\D$. Finally, in the same way we
prove that $S^{-1}\D_2$ is a field, so it is equal to the field of fractions $\K_2$ of the domain
$\D_2$.
\end{proof}

The next step is to rewrite our data according to \citep{Vas98}:

\begin{itemize}
    \item Let $f$ be the minimum common denominator of the rational
functions $f_i\in\K(\x)$.
    \item Let $\Phi:\ \K[y_1,\ldots,y_m]\to\K[\x,1/f]$,
defined as $\Phi(y_i)=f_i$ for each $i=1,\dots,m$.
    \item Let $\D_1=\Phi(\K[y_1,\ldots,y_m])=\K[f_1,\ldots,f_m]$.
We have that \linebreak $\D_1=\K[y_1,...,y_m]/\Ker(\Phi)$ is a
finitely generated $\K$-algebra. Also, the field of fractions of
$\D_1$ is $\K(f_1,...,f_m)$.
    \item Let $\D_2=\D_1[\x]=\K[\x,1/f]$.
The field of fractions of $\D_2$ is $\K(\x)$.
\end{itemize}

\subsection{Algorithm for the general case}

Summarizing the results we have presented, we have the following
algorithm to find intermediate unirational fields over a given
field, if the extension is separable.

\begin{algo}\label{alg-general} $ $
\begin{description}
    \item \textsc{Input}: $f_1,\ldots,f_m\in\K(\x)$.
    \item \textsc{Output}: rational functions $h_1,\ldots,h_r$
such that
\[\K(f_1,\ldots,f_m)\varsubsetneq\K(h_1,\ldots,h_r)\varsubsetneq\K(\x).\]
\end{description}
\begin{description}
    \item \textsc{A}. Compute the algebraic closure of
$\K(f_1,\ldots,f_m)$ relative to $\K(\x)$ according to Subsection
\ref{subsect-clausura}.

    \item \textsc{B}. Find a separating basis of
$\K(f_1,\ldots,f_m)$ according to Subsection
\ref{subsect-steinwandt}.

    \item \textsc{C}. Rewrite the fields according to Diagram
\ref{diagram}.

    \item \textsc{D}. Factor the minimum polynomial obtained
in the algebraic extension.

    \item \textsc{E}. Compute the decomposition blocks that
correspond to the factors found before.

    \item \textsc{F}. If such a block exists, due to Theorem
\ref{bloque-a-field}, we compute an intermediate field.

    \item \textsc{G}. Recover the generators of the intermediate
field in terms of the variables $x_1,\ldots,x_n$.
\end{description}
\end{algo}

%%%DAVID Añado el siguiente parrafo antes del ejemplo.
The following simple example follows the previous algorithm, but also shows a new way in which
intermediate fields can be computed more efficiently in some cases.

\begin{example}
Let $\F=\Q(x^4,y^6)\subset\Q(x,y,z)$. We want to find intermediate
fields of transcendence degree 2.

First, we will prove that the algebraic closure of $\F$ in
$\Q(x,y,z)$ is $\Q(x,y)$. Indeed, it is clear that this field is
algebraic over $\F$; on the other hand, no element $f\in\Q(x,y,z)$
with $\deg_z\,f>0$ can be algebraic over $\F$, as we would have a
non-zero polynomial that involves $x,y,z$.

As the closure of $\F$ in $\Q(x,y,z)$ is a rational field, we can
easily find intermediate fields: we decompose the generators and
obtain $1,x^2,x^4,y^2,y^3,y^6$. Each of the fields $\Q(x^4,y^6,f)$
where $f$ is one of the previous functions, is an intermediate
algebraic field. Not all of them can be expressed in this way, for
example $\Q(x^4,y^6,x+y)$. But we can construct linear combinations
of those to find primitive elements, in the same way as in Theorem.
As there are finitely many fields, this method may be a way of
computing them efficiently.
\end{example}

\subsection{Dimension and transcendence degree}

Now we present another method that reduces the general case to the
algebraic case. This time we will make use of the following theorem,
see \citep{Nag93} and \citep{AGR99}.

\begin{theorem}\label{th-ojan}
Let $x_1,\ldots,x_n$ be algebraically independent over an infinite
field $\K$. If $\F$ is a unirational field with
$\K\subset\F\subset\K(x_1,\ldots,x_n)$, there exist
$y_1,\ldots,y_d$ algebraically independent over $\K$ such that
$\F\subset\K(y_1,\ldots,y_d)$, where $d=\td(\K/\F)$.
\end{theorem}

The following algorithm is based on the proof given in the cited
paper.

\begin{algo}\label{alg-ojan} $ $
\begin{description}
    \item \textsc{Input}: $f_1,\ldots,f_m\in\K(\x)$.
    \item \textsc{Output}: an injective homomorphism $\Phi:\
\K(f_1,\ldots,f_m)\ \rightarrow\ \K(x_{i_1},\ldots,x_{i_d})$ where $d=\td(\K(f_1,\ldots,f_m)/\K)$.
\end{description}
\begin{description}
    \item \textsc{A}. Compute functions $\overline{f}_1,\ldots,\overline{f}_m$ such that:
\begin{description}
    \item --- $\K(\overline{f}_1,\ldots,\overline{f}_m)=\K(f_1,\ldots,f_m)$.
    \item --- $\overline{f}_1,\ldots,\overline{f}_d$ are algebraically independent over $\K$.
    \item --- $\overline{f}_{d+1},\ldots,\overline{f}_m$ are integral over $\K[\overline{f}_1,\ldots,\overline{f}_d]$.
\end{description}
If $d=m$, return $\Phi=id$.

    \item \textsc{B}. Reorder $x_1,\ldots,x_n$ so that:
\begin{description}
    \item --- $x_{d+1},\ldots,x_n$ are algebraically independent over $\K(\overline{f}_1,\ldots,\overline{f}_d)$.
    \item --- $x_1,\ldots,x_d$ are algebraic over $\K(\overline{f}_1,\ldots,\overline{f}_d,x_{d+1},\ldots,x_n)$.
\end{description}

    \item \textsc{C}. For each $i\in\{1,\ldots,d\}$ let
\[P_i(\overline{f}_1,\ldots,\overline{f}_d,x_{d+1},\ldots,x_n)\in\K[\overline{f}_1,\ldots,\overline{f}_d,x_{d+1},\ldots,x_n,z]\]
be non-constant and such that
$P_i(\overline{f}_1,\ldots,\overline{f}_d,x_{d+1},\ldots,x_n,x_i)=0$. Let $f$ be a common
denominator for $\overline{f}_1,\ldots,\overline{f}_d$ and write
$P_i=\displaystyle\frac{\widetilde{P}_i}{f^{r_i}}$ for adequate $r_i$'s. Let $\nu=\max\{\deg\
\widetilde{P}_i,\ \deg\ f,\ n\}+1$.

    \item \textsc{D}. Let $\varphi$ be the monomorphism
\[\begin{array}{cccc}
\varphi: & \K(f_1,\ldots,f_m) & \longrightarrow & \K(x_1,\ldots,x_{n-1}) \\
 & f_i(x_1,\ldots,x_n) & \rightarrow &
 f_i(x_1,\ldots,x_{n-1},x_1^\nu) \\
\end{array}\]
Let $\Phi=\varphi\circ id$.

    \item \textsc{E}. If $m-1=d$ return $\Phi$ after undoing the
reorder of the variables. Otherwise, repeat steps \textsc{B} to
\textsc{E} for $\Phi(\overline{f}_1),\ldots,\Phi(\overline{f}_m)$.
\end{description}
\end{algo}

\begin{analysis}
Computing the elements in \textsc{A} can be done due to a constructive proof of Noether's
Normalization Lemma. For step \textsc{B} it suffices to use Corollary \ref{th-sweedler-tag}. About
the definition of $\varphi$, the conditions on $\nu$ being greater than $\deg\ f$, $m$ and each
$\deg\ \widetilde{P}_i$ ensure that the application is well defined and is a monomorphism.

It is clear that the functions $f_1,\ldots,f_m$ and
$\varphi(\overline{f}_1),\ldots,\varphi(\overline{f}_m)$ have the
same properties as in \textsc{A}.

Regarding the complexity of the algorithm, it is dominated by the computation of Gr\"obner bases
in \textsc{B}; if we work in a general $\K$-algebra instead of a rational field, the computation
of the transcendence degree according to Subsection \ref{subsect-steinwandt} also needs Gr\"obner
bases.
\end{analysis}

We have proved that for a certain $\nu$, the homomorphism
\[(x_1,\ldots,x_n)\to(x_1,\ldots,x_{n-1},x_1^\nu)\]
is a monomorphism when restricted to $\K(f_1,\ldots,f_m)$. Let's see that we can use this to find
intermediate fields.

\begin{theorem}
Assume $\car\ \K=0$. Let $f\in\K(\x)$ be algebraic over \linebreak
$\K(f_1,\ldots,f_m)$. Then the application $\Phi$ that appears in
Algorithm \ref{alg-ojan} is also a monomorphism when we extend it to
$\F$.
\end{theorem}

\begin{proof}
As the extension is separable, we can write $\F=\K(f_1,\ldots,f_m,f)$. Applying this algorithm to
this representation of $\K$, in step \textsc{A} we can take the same
$\overline{f}_1,\ldots,\overline{f}_d$ as for $\K(f_1,\ldots,f_m)$ and, as there exists
$g\in\K(f_1,\ldots,f_m)$ such that $hg$ is integral over
$\K[\overline{f}_1,\ldots,\overline{f}_d]$, we take $\overline{f}_{m+1}=hg$. It is clear then that
in steps \textsc{B} and \textsc{C} we can reorder the variables and take the same polynomials.
From this we deduce that the value of $\nu$ that we had for $\K(f_1,\ldots,f_m)$ in step
\textsc{D} is also good for $\F$, and the same application is a monomorphism when extended to
$\F$.
\end{proof}

Due to this result, it is enough to apply the algorithm to the
given field, then we will have an algebraic extension
$\K(\Phi(f_1),\ldots,\Phi(f_m))\subset\K(x_{i_1},\ldots,x_{i_d})$.
The problem lies in how to compute $\Phi^{-1}(\E)$ for an
intermediate field in this extension, as showed in the next
elementary example.

\begin{example}
Let $\Phi:\ \K(x,y,z)\to\K(x,y)$ defined as
\[\Phi(x)=x,\ \Phi(y)=y,\ \Phi(z)=x^5.\]

Let $f=y^2\in\K(x,y)$, then
\[\left\{\frac{x^{5n}}{z^n}y^2+(z-x^5)\cdot g:\ n\in\Z,g\in\K(x,y,z)\right\}\subset\Phi^{-1}(f).\]
\end{example}

As there can be infinitely many candidates to inverse image of an
element, we cannot directly check them all. To complete this
solution, we would have to find a way to choose an algebraic
inverse image over $\K(f_1,\ldots,f_m)$.

\subsection{An idea based on a theorem by Schicho}

Another possible method for reducing the problem to another one in
an algebraic extension is based on rewriting the extension as a
simple extension,
\[\begin{array}{l}
\K(f_1,\ldots,f_m)=\K(\widehat{f}_1,\ldots,\widehat{f}_t)(f), \\
\F=\K(\widehat{f}_1,\ldots,\widehat{f}_t)(h), \\
\K(\x)=\K(\widehat{f}_1,\ldots,\widehat{f}_t,\widehat{f}_{t+1},\ldots,\widehat{f}_n)(g), \\
\end{array}\]
where $\{\widehat{f}_1,\ldots,\widehat{f}_t\}$ is a transcendence
basis of $\K(f_1,\ldots,f_m)$ and
$\{\widehat{f}_1,\ldots,\widehat{f}_n\}$ is one of de $\K(\x)$.

If we denote $\E=\K(\widehat{f}_1,\ldots,\widehat{f}_t)$ and
$\{\widehat{f}_{t+1},\ldots,\widehat{f}_n\}=\{z_1,\ldots,z_k\}$, we
have the fields
\[\E(f)\subset\E(h)\subset\E(z_1,\ldots,z_k,g)\]
so we are in the transcendence degree one case, with the exception
of working in a field where the variables are not independent. The
transcendence degree has been studied previously, see \citep{GRS01}.

In order to solve this with these  techniques,
we would need to adapt Theorem 3 in \citep{Schi95} to the field
$\E(z_1,\ldots,z_k,g)$ in the following way:

%%%DAVID No me gusta como lo tenias pero si no te gusta "conjetura" o "problema" pues vale como quieras

\begin{conjecture}
Let $A=\K(\x)$ and $B=\K(\y)$ two $\K$-algebras. Let $f_1,h_1\in A$ and $f_2,h_2\in B$ be
non-constant rational functions. Then these statements are equivalent:
\begin{description}
    \item[(i)] There exists a rational function $g\in\K(t)$
such that $f_1=g(h_1)$ and $f_2=g(h_2)$.
    \item[(ii)] $h_1-h_2$ divides $f_1-f_2$ in $A\otimes_\K B$.
\end{description}
\end{conjecture}

\section{$\K$-algebras}

Lastly, in this section we will briefly comment how to manipulate the elements involved from a computational point of
view when we work in a field of type $QF\left(\K[\x]/I\right)$ for some prime ideal $I \subset \K[\x]$ that is given
explicitly by means of a finite set of generators $I$.

The following known result asserts that any subfield in a finite extension is finitely generated.
A proof for zero characteristic fields is due to E. Noether, \cite{No}.

\begin{theorem}\label{teo-generated}
Let $\K\subset \K(z_1,\dots,z_n)$ be a finite extension. If $\F$ is a field such that
$\K\varsubsetneq \F\subset \K(z_1,\dots,z_n) $, then there exist $h_1,\dots,h_s\in
\K(z_1,\dots,z_n)$ such that $\F = \K(h_1,\dots,h_s)$.
\end{theorem}

As in previous sections, all the decision problems and computation
of the transcendence degree can be done for $\K$-algebras, see Theorem
\ref{th-sweedler-tag-Kalgebras}.

On the other hand, as the extension $\K\subset QF\left(\K[\x]/I\right)$ is not transcendental in
general, we need to ask that it is separable. We also can adapt Subsection
\ref{subsect-steinwandt} to this situation. Basically, we need to add the system of generators of
the ideal $I$ to $C$ in Theorem \ref{weil-t2} and Corollary \ref{weil-t3}. We illustrate this with
the following example.

\begin{example}\label{ej-steinwandt2}
We will work in the following field, which has transcendence degree
2 over $\Q$:
\[\Q(x,y,z)=QF(\Q[X,Y,Z]/(X^2+Y^2)).\]

Let $f_1=(x+2y-z)^3$, $f_2=(x+2y-z)^2$ in $\Q(x,y,z)$. We will
compute the transcendence degree of $\Q(f_1,f_2)$ over $\Q$.

A set of generators of the extended ideal is:
\[\begin{array}{l}
\{F_1=(X+2Y-Z)^3-(x+2y-z)^3, \\
F_2=(X+2Y-Z)^2-(x+2y-z)^2, \\
P=X^2+Y^2\}.
\end{array}\]

Deriving with respect to $X,Y,Z$ and evaluating in $x,y,z$ we obtain
\[\left(\begin{array}{ccc}
3\,(x+2\,y-z)^2 & 6\,(x+2\,y-z)^2 & -3\,(x+2\,y-z)^2\\
2\,x+4\,y-2\,z & 4\,x+8\,y-4\,z & -2\,x-4\,y+2\,z\\
2\,x & 2\,y & 0
\end{array}\right)\]

After some operations (remember that we are working in a $\Q$-algebra, so we must check that any
element we want to divide by is not zero, that is, it is not in the ideal of relations) we reach
an equivalent matrix:
\[\left(\begin{array}{ccc}
0 & 0 & -3(x+2\,y-z)^2\\
0 & 0 & -2(x+2\,y-z)^2\\
2\,x & 2\,y & 0
\end{array}\right)\]

It has rank 2, so $\td(\Q(x,y,z)/\Q(f_1,f_2))$ and $\td(\Q(f_1,f_2)/\Q)$ are both 1. Also, the
element $x$ and the element $y$ are transcendence bases of $\Q(x,y,z)$ over $\Q(f_1,f_2)$.
\end{example}

Also in Subsection \ref{subsect-clausura} we work in a setting that is general enough.

Once we reduce the problem to the algebraic case, we must consider
the rest of the algorithm. If we want to use the techniques
developed in Section \ref{sect-gradon} we must first ask that the
extension $\K(\x)/\K(f_1,\ldots,f_m)$ is separable.

\begin{remark}
It is enough that $\K(\x)/\K$ is separable. Indeed, then for each
intermediate field $\F$ there exists a separating basis $B$ such
that $\K(B)\subset\F$ is algebraic separable; then we only have to
find the fields in $\K(\x)$ and algebraic over $\K(B)$, and decide
which ones contain $\F$. To this end we will use Theorem
\ref{th-sweedler-tag-Kalgebras} to decide if the primitive element
for each field is in $\F$.
\end{remark}

About factorization in algebraic extensions and decomposition
blocks, we can work in a $\K$-algebra in the same way as a
rational field. However, the complexity increases dramatically,
because of two reasons: we must manipulate the representations of
the elements; and all the checks of type $f=0$ are transformed
into membership problems, $f\in\mathcal{B}_{\K(\x)/\K}$, that
demand Gr\"obner bases computations.

\section{Conclusions}

We have presented algorithms for resolving several issues related to rational function field. Our
approach has combined useful computational  algebra tools. Many interesting questions remain
unsolved. Unfortunately, we do not know if the computed intermediate field is rational or not. The
reason is that the algorithm produce an intermediate field generated always by the transcendence
degree plus one elements. It should be interesting to investigate under which circumstances our
algorithm can display an intermediate subfield generated by as many elements as the transcendence
degree. From a more practical point of view, we would like to have either a good algorithm or a
good implementation to compute a factorization of a polynomial over an algebraic extension.
Concerning applications, we suggest the possible use of  our techniques in the factorization of
morphisms and regular maps between affine and projective algebraic sets.


\begin{thebibliography}{}

\bibitem[Alonso, Gutierrez and Recio(1995)]{AGR95} C. Alonso,
J. Guti\'errez, T. Recio, \textsl{A rational function decomposition
algorithm by near-separated polynomials}. J. Symbolic. Comput. 19
(1995), no. 6, 527--544.

\bibitem[Alonso, Gutierrez and Rubio(1999)]{AGR99} C. Alonso,
J. Guti\'errez, R. Rubio, \textsl{On the dimension and the number of
parameters of a unirational variety}. Proceedings of CCNT'99,
Singapore, 3--9, Progr. Comput. Sci. Appl. Logic, 20, Birkh\"auser,
Basel, 2001.

\bibitem[Atiyah and MacDonald(1969)]{AMc69} M. F. Atiyah, I. G.
MacDonald, \textsl{Introduction to commutative algebra}. Addison
Wesley, 1969.

\bibitem[Becker and Weispfenning(1993)]{BW93} T. Becker, V. Weispfenning, \textsl{Groebner bases. A computational approach to commutative algebra }. In cooperation with Heinz Kredel. Graduate Texts in Mathematics, 141. Springer-Verlag, New York,  1993.

\bibitem[Brennan and Vasconcelos(1993)]{BV93} J. Brennan,
W. Vasconcelos, \textsl{Effective computation of the integral
closure of a morphism}. J. Pure Appl. Algebra 86 (1993), no. 2,
125--134.

\bibitem[Casperson, Ford and McKay(1996)]{CFM96} D. Casperson, D. Ford, and J. McKay. \textsl{Ideal
decompositions and subfields}. J. Symbolic Comput. 21 (1996), no. 2,
133--137.

\bibitem[von zur Gathen and Gerhard(1999)]{GG99} J. von zur Gathen,
J. Gerhard, \textsl{Modern Computer Algebra}. Cambridge University
Press, New York, 1999.

\bibitem[Gianni and Trager(1997)]{GT97} P. Gianni, B. Trager, \textsl{Integral
closure of Noetherian rings}. Proceedings of ISSAC'97 (Kihei, HI),
212--216 (electronic), ACM Press, New York, 1997.

\bibitem[Gutierrez, Rubio and Sevilla(2001)]{GRS01} J.
Guti\'errez, R. Rubio, D. Sevilla, \textsl{Unirational fields of
transcendence degree one and functional decomposition}. ISSAC
2001, London, Canada, 167--174.

\bibitem[Gutierrez, Rubio and Sevilla(2002)]{GRS02}
J. Guti\'errez, R. Rubio, D. Sevilla, \textsl{On multivariate
rational function decomposition}. Computer algebra (London, ON,
2001). J. Symbolic Comput. 33 (2002), no. 5, 545--562.


\bibitem[Hommel and Kov\'acs(1992)]{HK92}
G. Hommel, P. Kov\'acs,
\textsl{Simplification of symbolic
inverse kinematic transformations through functional decomposition}.
Proc. of the Conference Adv. in Robotics, Ferrara, 88--95
(1992).\par


\bibitem[Kl\"uners and Pohst(1997)]{KP97} J. Kl\"uners, M. Pohst,
\textsl{On computing Subfields}. J. of Symbolic Computation, 24 (1997),
385--397.

\bibitem[Landau(1985)]{Lan85} S. Landau, \textsl{Factoring polynomials
over algebraic number fields}. SIAM J. Comput. 14 (1985), no. 1,
184--195.

\bibitem[Landau and Miller(1985)]{LM85} S. Landau, G. L. Miller,
\textsl{Solvability by radicals is in polynomial time}. J. Comput.
System Sci. 30 (1985), no. 2, 179--208.

\bibitem[Lang(1967)]{Lan67} Lang, S.\textsl{ Algebra.}
Addison--Wesley, Reading, Mass (1967).

\bibitem[Lazard and Valibouze(1993)]{LV93} D. Lazard, A. Valibouze,
\textsl{Computing subfields: reverse of the primitive element
problem}. Computational algebraic geometry (Nice, 1992), 163--176,
Progr. Math., 109, Birkh\"auser Boston, Boston, MA, 1993.

\bibitem[Lenstra, Lenstra and Lov\'asz(1982)]{LLL82} A. K. Lenstra,
H. W. Lenstra, L. Lov\'asz, \textsl{Factoring polynomials with
rational coefficients}. Math. Ann. 261 (1982), no. 4, 515--534.

\bibitem[Loos(1983)]{Loo83} Loos, R. \textsl{Computing in algebraic
extensions}. Computer algebra, Springer, Vienna, 1983.

\bibitem[M\"uller-Quade and Steinwandt(1999)]{MS99} J. M\"uller-Quade, R. Steinwandt, \textsl{Basic
algorithms for rational function fields}. J. Symbolic Comput. 27
(1999), no. 2, 143--170.

\bibitem[Nagata(1993)]{Nag93} M. Nagata, \textsl{Theory of
commutative fields}. Translations of Mathematical Monographs, 125.
American Mathematical Society, Providence, RI, 1993.

\bibitem[Noether(1915)]{No} E. Noether,
 \textsl{K\"orper und Systeme rationaler Funktionen}.
 Math. Ann. {\bf 76}, 161--196 (1915).\par

\bibitem[Rubio(2001)]{Rub01} R. Rubio, \textsl{
Unirational fields. Theorems, algorithms and
applications}. PhD. Thesis.
 Dep. of Mathematics, University
of Cantabria, Spain, 2001


\bibitem[Shafarevich(1977)]{Sha77} I.R. Shafarevich, \textsl{
Basic Algebraic Geometry}.
 Springer Study Edition,
Springer-Verlag, 1977.



\bibitem[Schicho(1995)]{Schi95} J. Schicho, \textsl{A note on a theorem of Fried and MacRae}. Arch. Math. 65, 239-243,
1995.

\bibitem[Schinzel(1982)]{Sch82} A. Schinzel, \textsl{Selected
topics on polynomials}. Ann Arbor, University of Michigan Press,
1982.

\bibitem[Seidenberg(1975)]{Sei75} A. Seidenberg, \textsl{Construction
of the integral closure of a finite integral domain. II}. Proc.
Amer. Math. Soc. 52 (1975), 368--372.

\bibitem[Steinwandt(2000)]{Ste00} R. Steinwandt, \textsl{On computing a
separating transcendence basis}. SIGSAM Bulletin, 34(4): 3-6,
2000.

\bibitem[Sweedler(1993)]{Swe93} M. Sweedler, \textsl{Using Gr\"obner
bases to determine the algebraic and transcendental nature of
field extensions: return of the killer tag variables}. Applied
algebra, algebraic algorithms and error-correcting codes (San
Juan, PR, 1993), 66--75, Lecture Notes in Comput. Sci., 673,
Springer, Berlin, 1993.

\bibitem[Trager(1976)]{Tra76} B. Trager, \textsl{Algebraic factoring and
rational function integration}. Proc. 1976 ACM Symp. Symbolic 6
Algebraic Comp., 219--228, 1976.

\bibitem[Vasconcelos(1998)]{Vas98} W. Vasconcelos, \textsl{Computational
Methods in Commutative Algebra and Algebraic Geometry}. Vol. 2 of
Algorithms and Computation in Mathematics, Springer-Verlag, 1998.

\bibitem[van der Waerden(1964)]{Wae64} B. L. van der Waerden,
\textsl{Modern Algebra}. Frederick Ungar Publishing Co., New York,
1964.

\bibitem[Weil(1946)]{Wei46} A. Weil, \textsl{Foundations of Algebraic
Geometry}. American Mathematical Society Colloquium Publications,
vol. 29. American Mathematical Society, New York, 1946.

\bibitem[Wielandt(1964)]{Wie64} H. Wielandt, \textsl{Finite permutation
groups}. Academic Press, New York, London, 1964.

\bibitem[Yokoyama, Noro and Takeshima(1989)]{YNT89} K. Yokoyama, M. Noro, T. Takeshima,
\textsl{Computing primitive elements of extensions fields}. J.
Symbolic Comput. 8 (1989), no. 6, 553--580.

\bibitem[Zippel(1991)]{Zip91} R. Zippel, \textsl{Rational function
decomposition}. Proc. ISSAC'91, ACM press, 1991.

\bibitem[Zippel(1993)]{Zip93} R. Zippel, \textsl{Effective
polynomial computation}. Kluwer Academic Press, 1993.


\end{thebibliography}
\end{document}